# The influence of sonication on the thermal behavior of muscovite and biotite

Luis A. Pérez-Maqueda[a,]*, José M. Blanes[a], José Pascual[b], José L. Pérez-Rodríguez[a]

[a]*Instituto de Ciencia de Materiales de Sevilla, CSIC, Universidad de Sevilla, Américo Vespucio s/n 41092 Sevilla, Spain*
[b]*Departamento de Ingeniería Civil, Materiales y Fabricación, ETSII, Universidad de Málaga, Plaza, Ejido s/n 29013 Málaga, Spain*

## Abstract

The differences on the thermal behavior (DTA-TG) of mica samples measured before and after sonication have been studied. Sonication treatment produces important modifications in the thermal behavior of muscovite and biotite samples. For muscovite, it produces a broadening and decrease in temperature of the dehydroxylation and crystallization effects, reaching a steady stage after 40 h treatment time. For biotite, the original single peak profile for the dehydroxylation of the untreated sample is converted into a two peaks profile after sonication, the intensity of the low temperature peak increases with sonication time, while the intensity of the high temperature peak decreases. The modification of the thermal behavior for sonicated samples has been correlated to the particle size distribution effect produced by the sonication treatment. It has been also observed that the cup tip of the sonication equipment contaminates the samples releasing titanium of its composition.

## 1. Introduction

Micas are minerals of significant commercial importance due to their physical properties. These minerals remain stable at high temperature and have interesting electrical properties. Therefore, they are used in applications where it is required high-temperature stability and electrical properties.[1]

The thermal behavior of micas has been extensively studied. Thus, studies on the dehydroxylation of dioctahedral micas, such as celadonite and glauconite, have shown that during dehydroxylation cations migrate from *cis* into *trans*-octahedra and have five-fold coordination.[2] For muscovite, Mackenzie et al.[3] have suggested an homogeneous dehydroxylation mechanism forming a dehydroxylate in which the aluminum is predominantly 5-coordinate. On the other hand Guggenheim et al.[4] have found a very broad peak in the differential thermogravimetric (DTG) trace of a mica sample that they interpreted as two overlapping poorly resolved dehydroxylation peaks. These two peaks have been explained by considering that the Al–OH bond is greatly affected by the coordination number of neighboring polyhedron. When neighboring polyhedron are in octahedral coordination, the hydroxyl group is lost at lower temperatures than when neighboring polyhedron are in five-fold coordination (after partial dehydroxylation).

Mechanical treatment are of great importance in the preparation and processing of raw materials.[5] Micas are often used in ground forms. Mica may be dry ground (yielding particles in the range from 1.2 mm to 150 mm), wet ground (95–45 mm), or micronized (< 53 mm). When studying grinding treatments emphasis has been placed on the impact of milling on the physicochemical properties of the initially coarse powder, and its influence on the transformation of the solid.[6,7] It is well known that grinding of clay minerals produces various effects on their structure and properties.[8] The significant processes involved in the preparation of ceramic raw materials have been extensively studied, specially those for kaolinite,[9,10] montmorillonite or bentonite,[11] illite,[12] pyrophyllite,[13] talc,[14] and vermiculite.[15]

It has been observed that grinding of clays produces progressive amorphyzation when grinding time increa-ses.[16] In many cases this treatment yields heterogeneous hardly aggregated materials with modified chemical reactivity. All these effects are useful for some processes, but they are important drawbacks for many of the applications of mica, where it is important to have small particle diameters while maintaining the crystalline structure in order to retain the properties of mica. An alternative method for particle size reduction is sonication. Cavitational collapse of bubbles on solid surfaces leads to micro jet and shock-wave impacts on the surface of the solids, together with interparticle collisions which can result in particle size reduction.[17,18] Recently, micron and submicron-sized vermiculite flakes were prepared from a natural macroscopic vermiculite sample by sonication.[17,19] The resulting materials were crystalline, as assayed by X-ray diffraction.[20]

---

* Corresponding author.
  *E-mail address:* maqueda@cica.es (L.A. Pérez-Maqueda).

Particle size distribution is a very sensitive issue in the thermal behavior of materials. The importance of this issue has been recognized in the literature and authors have observed changes in the profiles of the thermal transformation of solid when performing experiments under both isothermal and non-isothermal conditions.[21–24] For kinetically driven process, it has been shown in a theoretical work[22] that the shape and average temperature of the thermogravimetric curves are affected by the particle size distribution. This change does not necessarily imply a modification in the kinetic parameters (activation energy, pre-exponential parameter of Arrhenius, kinetic model) of the reaction.

Clay minerals subjected to grinding experience particle size reduction together with other important transformations, such as amorphyzation, aggregation of small particles into larger units and contamination. The influence of grinding on the thermal behavior of minerals has been extensively studied.[16,25–28] However, until now, no consideration has been given to the effect of sonication on the thermal behavior of minerals. The purpose of this work is to show the effect of sonication on the thermal behavior of muscovite and biotite, special attention is given to the effect of particle size distribution. In addition, the influence of the possible contamination induced by the tip cups of the ultrasound equipment on the thermal analysis and new phases produced is investigated.

2. Experimental

2.1. Materials

A muscovite from Fuente Obejuna (Córdoba, Spain) and a biotite from Santa Olalla (Huelva, Spain) were used as starting materials.[29] Both samples consist of platelets of about 10 cm in length and 0.5 cm in thickness. Previously to the sonication treatment, samples were lightly ground using a knife-mill and sieved under 2 mm.

2.2. Sonication

A high-intensity ultrasonic horn (Misonix inc.), that consists of a solid titanium rod connected to a piezoelectric ceramic, and a 20 KHz, 750 W power supply were used. The horn tip was introduced into a thermostated (20 °C) double-jacket reactor. Three grams of < 2 mm flakes were mixed with 100 cm$^3$ of freshly deionized water and subjected to ultrasounds for periods ranging between 10 and 100 h.

2.3. X-ray diffraction analysis

Diffraction patterns were obtained using a diffractometer (Kristalloflex D-500 Siemens) at 36 kV and 26 mA with Ni-filtered Cu$K\alpha$ radiation and a graphite monochromator.

2.4. Nuclear magnetic resonance measurements

High-resolution solid-state $^{29}$Si and $^{27}$Al magic angle spinning (MAS) nuclear magnetic resonance (NMR) spectra of powdered samples were recorded at 79.49 and 104.26 MHz, respectively, in a Brüker MSL-400 spectrometer. Measurements were conducted at room temperature with tetramethylsilane (TMS) and [Al(H$_2$O)6]$^{3+}$ as external references.

2.5. Thermal study

Thermogravimetric analysis (TG), differential thermogravimetric analysis (DTG) and differential thermal analysis (DTA) were carried out simultaneously in static air or argon flow (500 cm$^3$ min$^{-1}$) with an automatic thermal analyzer system (Seiko, TG/DTA 6300). Mica (muscovite and biotite) samples of about 50 mg were loosely packed into a platinum holder and were thermally treated at a heating rate of 10 °C min$^{-1}$.

2.6. Scanning electron microscope study

The samples were studied by scanning electron microscopy (SEM) using a Jeol JSM-5400 model. The samples were covered with a thin gold film and analyzed by energy dispersive X-ray spectroscopy (EDX) using a Link-ISIS Si/Li detector. The mean size (length, L) of the plate-like particles was evaluated by measuring the length in the longest direction of around 100 particles for each sample.

2.7. Particle size distribution

The mass fraction versus particle diameter plots have been obtained by a deposition–centrifugation separation procedure described in the literature[30] followed by the gravimetric quantification of the different fractions.

*2.8. Chemical analysis*

Chemical quantification of Ti in bulk samples was carried out using a sequential X-ray fluorescence spectrometer (Siemens, SRS 3000).

3. Results

The $^{27}$Al and $^{29}$Si NMR analysis of the untreated and sonicated muscovite samples (figures not shown) did not show at the atomic scale any modification in the coordination of aluminum or silicon.[29] Thus, for both sonicated and untreated samples, the $^{27}$Al spectra showed the typical bands corresponding to aluminum in octahedral and tetrahedral sites[3,31] and the $^{29}$Si spectra showed a peak attributed to Si(Si$_2$Al) environment.[3,32] Additionally, the X-ray diffraction patterns of the original and sonicated muscovite and biotite samples indicated that sonication does not produce significant structural changes in the samples. Thus, the diffraction lines of both untreated and sonicated samples remained unchanged, except for the broadening of the lines, that could be attributed to the delamination and crystallite size reduction,[17] and to the change in the relative intensity of the diffractions due to textural effects.[33] These results indicate that the structure of the micas is maintained after the sonication treatment in contrast with those results previously obtained for ground clay minerals,[9,13,16,25—27] where even short milling treatments produce significant structure modifications.

The gradual size reduction on sonication of muscovite and biotite was revealed by SEM (Fig. 1). During sonication treatment, the original stacking layers of muscovite and biotite are delaminated and the lamellar phyllosilicate mineral particles are broken by mechanical impact induced by sonication producing a decrease in particle size. The evolution of the particle length with sonication time (Fig. 2) shows for the muscovite an important decrease of particle length with sonication time up to a limit reached for 40 h. For the biotite, the particle length decreases in the entire sonication range studied here. The particle lengths of the sonicated samples are larger for biotite than for muscovite.

The mass fraction as a function of the particle diameter plots for the muscovite and biotite samples sonicated for 40 h are presented in Fig. 3. These plots show that the sonicated samples are not monodispersed but constituted of different particle size fractions in the range of micron and submicron sizes. It is also clear from this figure, as it was also revealed from SEM measurements, that the sonicated biotite sample has a larger particle size than the muscovite one.

DTG diagrams carried out in argon flow of untreated and sonicated muscovite samples in the range from 300 to 1000 °C, where dehydroxylation of muscovite takes place, are shown in Fig. 4. The weight loss of the untreated sample begins at 650 °C and reaction is over at 920 °C. The total weight loss agrees with the theoretical one calculated for an ideal muscovite (≈4.70%). This theoretical weight loss is assumed to be due entirely to the structural water lost by dehydroxylation as estimated from the crystal structure. Sonication modifies the DTG profile (Fig. 4). The relatively sharp weight loss centered at 824 °C for the untreated sample is converted into a much broader weight loss shifted to lower temperatures. Thus, the DTG maximum for the sample sonicated for 10 h (Fig. 4b) appears at 700 °C (125 °C lower than for the untreated sample). Longer sonication times produce even a more pronounced shifting of this DTG peak. This shifting effect reaches a limit for 40 h sonication time (maximum at 640 °C, Fig. 4c). In addition to this weight loss, a new weight loss is observed for the sonicated samples in the range from 300 to 450 °C that appears as a shoulder in the DTG trace (Fig. 4b–d). This weight loss also increases with sonication time up to a limit reached for 40 h treatment time.

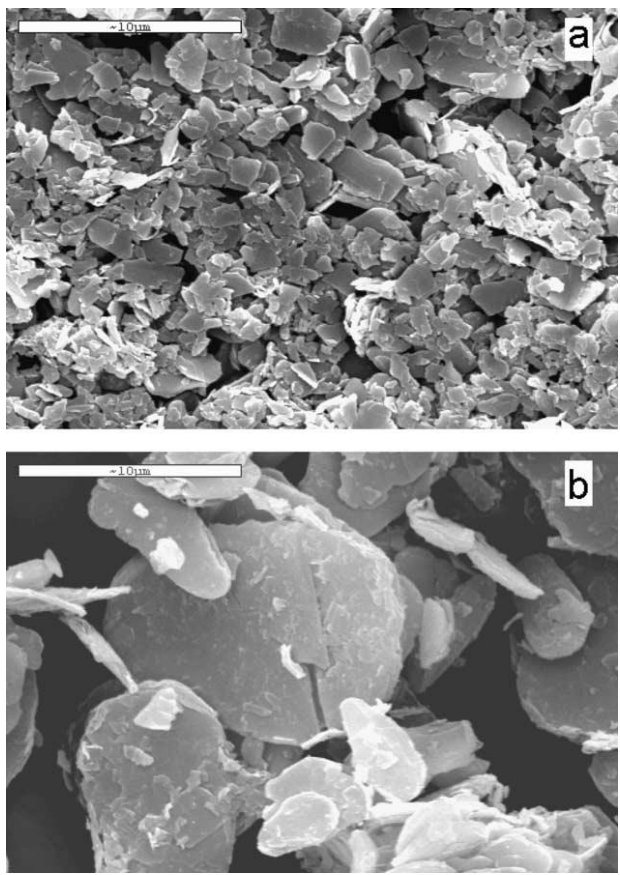

Fig. 1. Scanning electron micrographs of muscovite (a) and biotite (b) samples sonicated for 40 h.

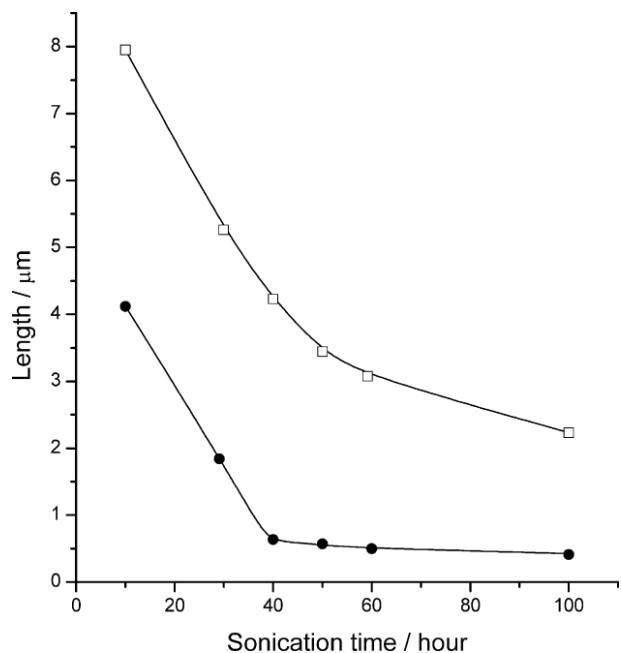

Fig. 2. Evolution of the mean size (length) as a function of the sonication time for biotite (&) and muscovite (*).

Since X-ray diffraction and NMR studies have shown that sonication does not produce significant changes in the mica crystalline structure or Al and Si coordination, the effect of sonication on the thermal behavior of the muscovite could be attributed to changes in the particle size distribution. The importance of particle size in the thermal reactivity of solids have been recognized in literature.[21—24] Theoretical studies[22] have shown that, for kinetically controlled processes, the shape of the TG-DTG traces and the average temperature of the process are affected by the particle size distribution. Thus, in two identical samples with different particle size distribution, the differences in their thermal behavior could be attributed to the particle size effect and not necessarily to a change in the kinetic mechanism or in the kinetic parameters (activation energy or pre-exponential factor of Arrhenius).

For the muscovite sample, there is a correlation between the effect of sonication on particle size reduction and the changes in the thermal behavior. Thus, the thermal behavior changes while the particle size is reduced, that is up to 40 h treatment time, remaining unchanged for longer treatments times. The particle size reduction produced by sonication facilitates the dehydroxylation of the muscovite as observed by the shifting of the main weight loss toward lower temperature. In addition, the new borders created by the particle size reduction should be responsible of the low temperature (300–450 °C) due to weakly bonded OH groups placed on these borders.

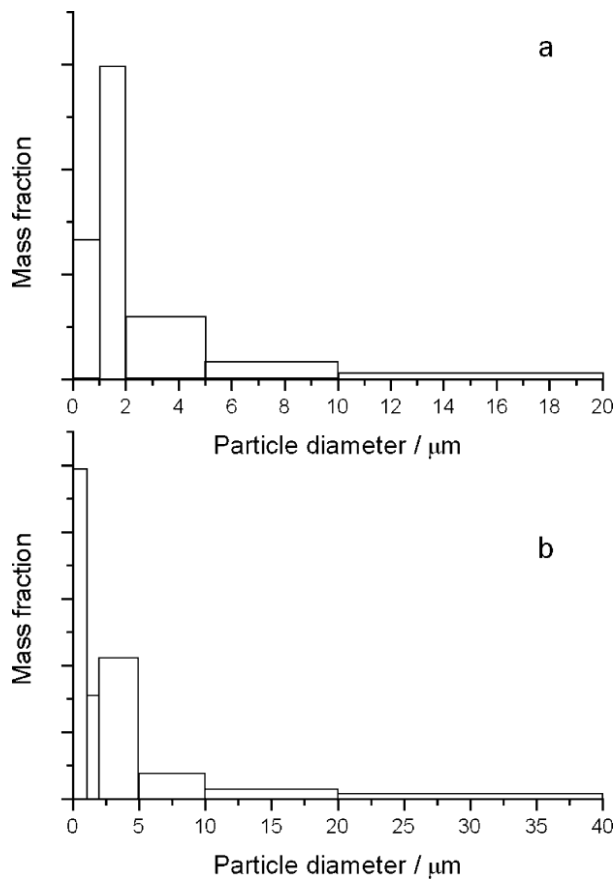

Fig. 3. Mass fraction as a function of the particle diameter for the muscovite (a) and biotite (b) samples sonicated for 40 h.

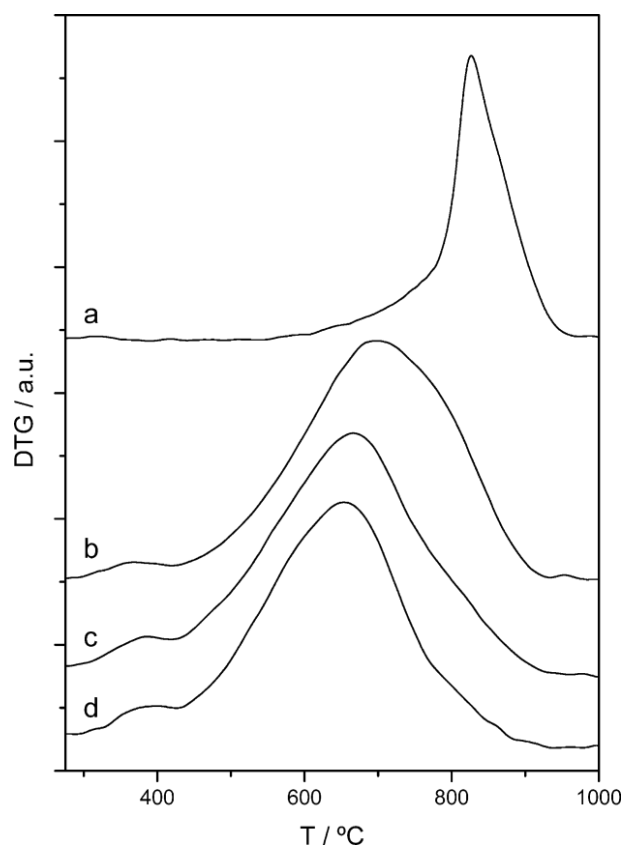

Fig. 4. DTG curves obtained under argon flow at a heating rate of 10 °C m$^{-1}$ for the muscovite before (a) and after sonication for 10 h (b), 40 h (c), and 100 h (d).

In order to study the influence of the particle size distribution on the thermal behavior of the sonicated muscovite, DTG plots of the different fractions with different particle sizes separated by the deposition–centrifugation procedure were recorded for the sample sonicated for 40 h under argon atmosphere. Fig. 5 shows the DTG traces for the different size fractions after normalization and multiplication by the amount of sample as taken from Fig. 3, the overall curve resulted of adding up of the DTG traces of the different fractions is also included (solid line). It is clear from this figure (Fig. 5) that the overall trace resulting of adding up the contribution of all the fractions has the same shape as the DTG trace shown in Fig. 4b for the muscovite sonicated for 40 h. Fig. 5 allows to discriminate the influence of the different particle sizes on the DTG trace. Thus, the fractions with smaller particle size have a more siginificant contribution to the low temperature weight loss than the larger particles. In addition, the main weight loss with a maximum in the DTG at 640 °C is the result of the contributions of all the fractions in such a way that the smaller particles dehydroxylate at lower temperature than the larger ones. The broadening of the DTG peak of the sonicated samples is also due to the contribution of the different particle size fractions. These results illustrate the effect that the changes in the mean particle size and particle size distribution have on the thermal dehydroxylation of muscovite.

The DTA curve of the original muscovite carried out in argon flow (Fig. 6a) shows two endothermic peaks, one with a minimum at 825 °C and the other with a minimum at 1145 °C. The temperature of former endothermic peak matches that of the weight loss in the DTG curve (Fig. 4a) and corresponds to dehydroxylation. The endothermic peak at higher temperatures represent recrystallization into leucite, corundum and mullite, as was confirmed by X-ray diffraction of the sonicated muscovite samples heated to 1250 °C (figure not shown). The products of recrystallization may vary somewhat for different natural muscovite samples.[34] In Fig. 6b, it has been included the DTA of the sample sonicated for 100 h. For this sample, the dehydroxylation effect is also broadened and shifted to lower temperatures as compared with that of the untreated material. The evolution of this peak is identical to that observed for the DTG trace (Fig. 4). The DTA effect at temperature higher than 900 °C suffers also broadening and shifting to lower temperature with the sonication time. The shifting of the higher temperature endothermic peak has also been observed for ground muscovite.[34]

Fig. 6c shows the DTA diagram recorded in static air for the sonicated muscovite sample. The endothermic effects described above for the experiments performed in argon are also observed in the DTA traces recorded in air. However, an additional exothermic effect at 682 °C is present in treated sample. TG experiments (figure not shown) correlate this exothermic effect with an increase in weight at the same temperatures. This exothermic effects and the corresponding weight increase may be only attributed to oxidation and/or nitridation of some compound present or formed during the sonication treatment. The chemical analysis as assayed by X-ray fluorescence of the muscovite and biotite samples sonicated for different times show similar composition, except for the percentages of $TiO_2$ that suffer a remarkable increase in relation with the untreated samples. Thus, the $TiO_2$ percentages change from 0.05% for untreated sample to 5.33% in muscovite sonicated for 100 h. This increase in titanium is due to the release of this element from the tip cup of the ultrasound equipment that is manufactured of Ti. The powder X-ray diffraction studied of sonicated muscovite and biotite samples confirms the presence of titanium element that is oxidized to titanium oxide, rutile, during the heating in static air (figure not shown). The muscovite sample after 100 h of sonication and heated at 1200 °C (Fig. 7A) is constituted by platelets particles constituted by Si, Al and K (Fig. 7B) that correspond to the muscovite material and spherical particles constituted by titanium oxide (Fig. 7C). These data confirm that sonicated samples are contaminated by particles of Ti that is release from the tip cup of the ultrasound probe. This Ti is present as discrete particles mixed with the mica particles. The Ti particles are oxidized to rutile when heated under static air remaining as independent particles mixed with the mica. While the importance of contamination during grinding of minerals has been previously recognized in the literature,[25,27] no data are published, up to our knowledge, on the contamination of solids subjected to sonication treatment.

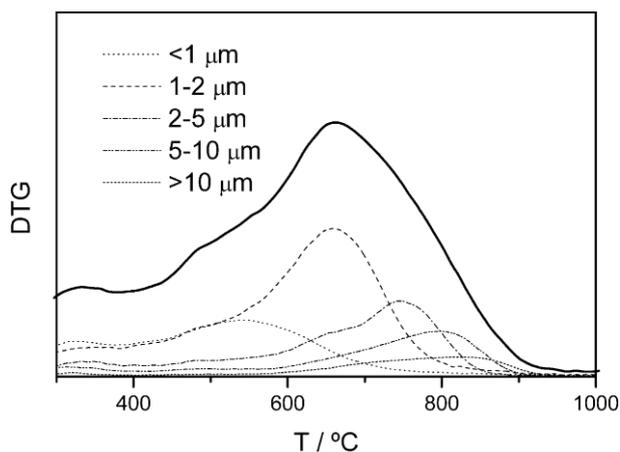

Fig. 5. DTG traces of the different fractions with different particle sizes separated by the deposition–centrifugation procedure recorded for the muscovite sample sonicated for 40 h under argon flow at a heating rate of 10 °C m$^{-1}$. The overall curve resulted of adding the DTG traces corresponding to the different fractions is plotted as a

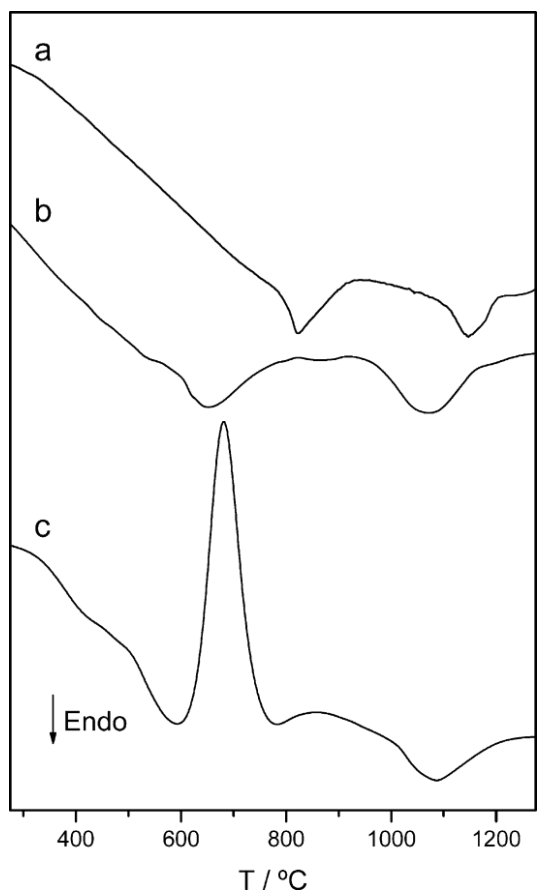

Fig. 6. DTA curves at a heating rate of 10 °C min$^{-1}$ for the untreated muscovite in argon flow (a) and for the muscovite sonicated for 100 h in argon flow (b) and in static air (c).

For biotite, sonication has also a significant effect on its thermal behavior. the DTG curves recorded in argon flow in the range 900–1225 °C of biotite samples before and after sonication are shown in Fig. 8. This temperature range has been selected because the thermal dehydroxylation of biotite takes place in this range. The DTG curve for the untreated biotite (Fig. 8a) shows a peak with a maximum at 1160 °C. The weight loss corresponding to dehydroxylation, as would be expected for a magnesium containing mineral, takes place at a considerably higher temperature than that of an aluminum containing mineral muscovite. After 10 h sonica-tion time, the DTG trace (Fig. 8b) shows that the peak centered at 1160 °C is broader and it is also accompanied by a shoulder at lower temperature. This shoulder is transformed into a peak at 1052 °C after 20 h treatment (Fig. 8c). The intensity of this second peak increases with sonication time, while the intensity of the high temperature peak decreases. Thus, after 100 h treatment time the high temperature peak has almost disappeared and it is only present as a small shoulder (Fig. 8f).

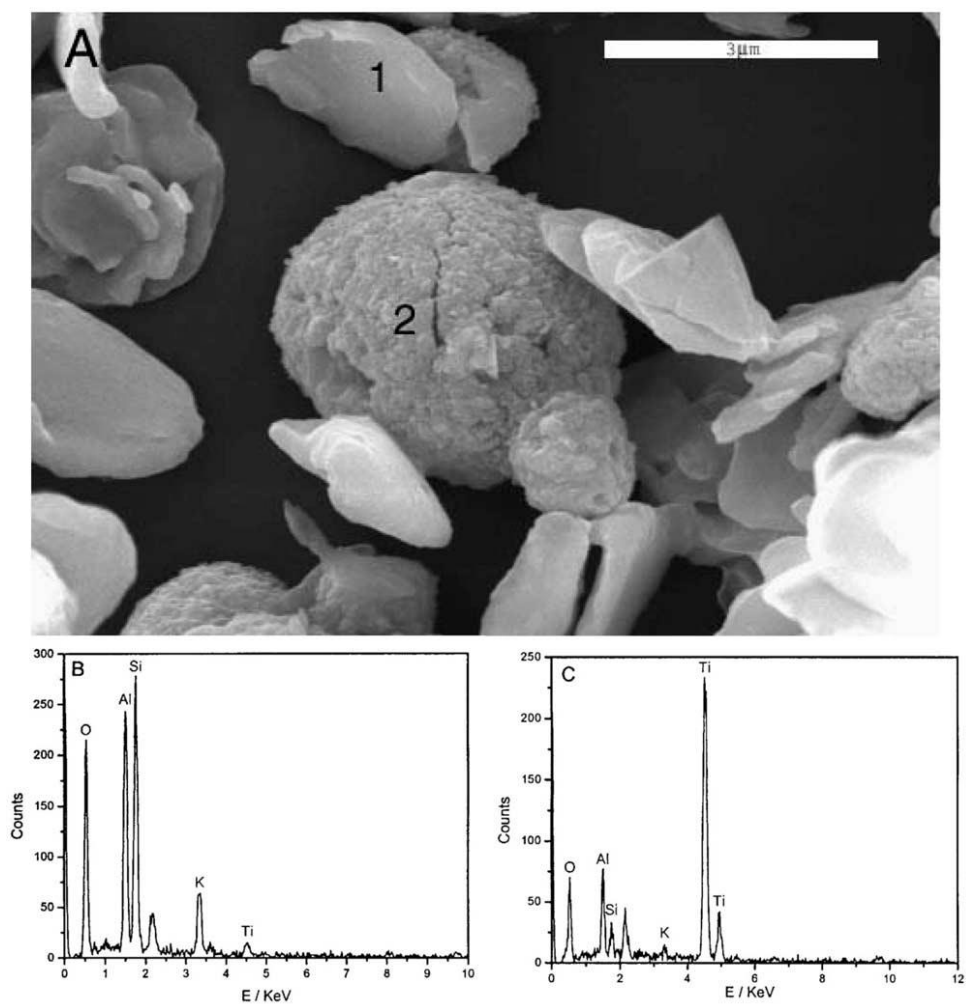

Fig. 7. (A) Scanning electron micrograph of the muscovite sample sonicated for 100 h and heated at 1200 °C. (B) EDX analysis of platelets particles marked in the micrograph as 1. (C) EDX analysis of spherical particles marked in the micrograph as 2.

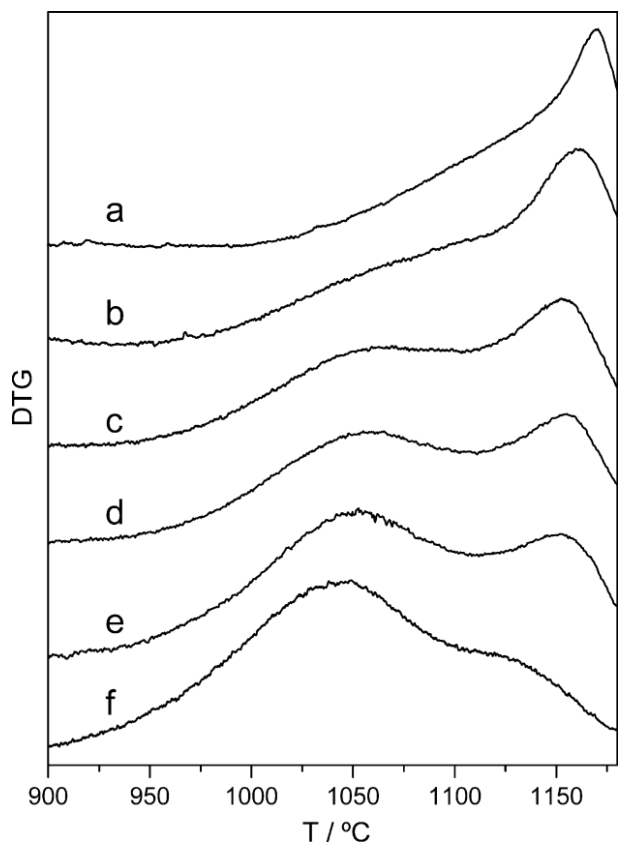

Fig. 8. DTG curves obtained under argon flow and a heating rate of 10 °C min$^{-1}$ for the biotite sample before (a) and after sonication for 10 h (b), 20 h (c), 40 h (d), 50 h (e) and 100 h (f).

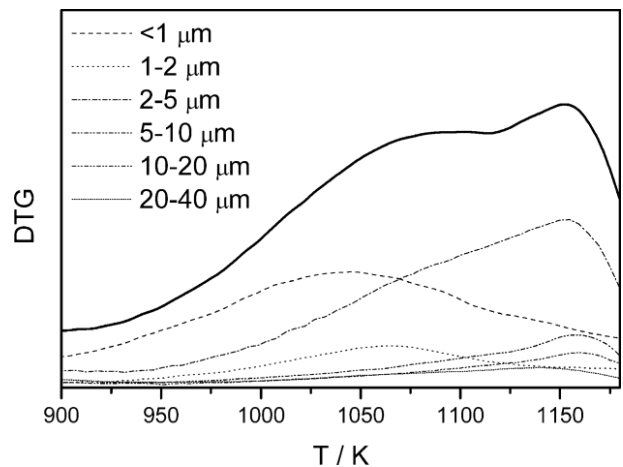

Fig. 9. DTG traces of the different fractions with different particle sizes separated by the deposition-centrifugation procedure recorded for the biotite sample sonicated for 40 h under argon flow at a heating rate of 10 °C m$^{-1}$. The overall curve resulted of adding the DTG traces corresponding to the different fractions is plotted as a solid line.

In order to study the contribution of the different particle sizes on the thermal behavior of the biotite sample, the same particle size separation procedure used for muscovite was applied to biotite. Fig. 9 includes for the biotite sample sonicated for 40 h, the DTG trace of the different fractions with the overall curve resulted of adding up the contribution of all the fractions. In this case, as it happened for muscovite, the curve resulted of adding up all the individual DTG traces present the same shape as that included in Fig. 8d for the sample sonicated for 40 h. Fig. 9 shows that particles larger than 2 mm dehydroxylate at higher temperatures than those smaller than 2 mm. Thus, the DTG peak at higher temperature (with a maximum at 1160 °C) is due to the dehydroxylation of particles with an average size larger than 2 mm, while the lower temperature DTG peak (with a maximum at 1050 °C) is mostly due to particles smaller than 2 mm. For biotite, unlike muscovite where particle size reduction is leveled off at 40 h sonication time, the particle size reduction takes place along the entire sonication time. This continuous particle size reduction has an effect also in the thermal behavior. Thus, the lower temperature peak increases while the high temperature effect decreases with sonication time, indicating that the amount of particles with sizes smaller than 2 mm is increasing while the amount of the larger ones is decreasing.

The DTA and TG performed in air for the sonicated biotite samples (figures not shown) also indicated contamination of Ti that oxidizes during heating. This Ti contamination was also detected by X-ray diffraction, chemical analysis and SEM in a similar way as in the muscovite samples.

4. Conclusions

Sonication does not produce significant structural transformation in the mica samples. Additionally, for the muscovite sample, it has been observed by NMR that the coordination of Al and Si does not suffer modification due to the treatment. Nevertheless, the particle size of the samples is drastically reduced. This particle size reduction produces important changes in the thermal behavior of these micas. For the muscovite sample, the weight losses on the DTG curve due to dehydroxylation and the corresponding DTA endothermic effect are broadened and shifted to lower temperature, until a sonication time limit is reached at 40 h, treatments longer than 40 h do not produce any impor-tant change in the TG or DTA traces. This behavior could be correlated with the decrease of particle size up to 40 h sonication time followed by a leveling off. The broadening of the main weight loss is related to the effect of the particle size distribution. The weight loss in the range from 300 to 450 °C could be attributed to loosely OH in the borders of the particles generated by the sonication treatment.

For the biotite, the particle size decreases in the entire sonication time range studied here, and this continuous change has an important effect on the biotite thermal behavior. Thus, the only weight loss in the range from 900 to 1200 °C due to dehydroxylation observed for the untreated biotite is converted into two different weight losses for the sonicated samples. The DTG trace shows that one of the peak increases its intensity while the other decreases with the sonication time. The study of the influence of particle size distribution on the thermal behavior has shown that the low temperature peak is due to particles smaller than 2 mm while the one at higher temperatures is due to particles larger than 2 mm.

The contamination of sonicated samples has been reported, up to our knowledge, for the first time in this paper. Thus, it has been observed that Ti is released from the sonication probe and appears as individual particles mixed with the sonicated particles.

Acknowledgements

This publication was prepared on the basis of the results obtained in the frame of the project MAT 2002-03774 of the Ministry of Science and Technology of Spain.

References

1. Hedrick, J. B., *Mica. Am. Ceram. Soc. Bull.*, 1999, 78, 136–138.
2. Muller, F., Drits, V., Plancon, A. and Besson, G., Dehydroxylation of $Fe^{3+}$, Mg-rich dioctahedral micas: (I) structural transformation. *Clay Miner.*, 2000, 35, 491–504.
3. Mackenzie, K. J. D., Brown, I. W. M., Cardile, C. M. and Meinhold, R. H., The thermal reactions of muscovite studied by high-resolution solid-state $^{29}Si$ and $^{27}Al$ NMR. *J. Mater. Sci.*, 1987, 22, 2645–2654.
4. Guggenheim, S., Chang, Y.-H. and Koster van Groos, A. F., Muscovite dehydroxilation: high-temperature studies. *Am. Mineralogist*, 1987, 72, 537–550.
5. Ovadyahu, D., Yariv, S., Lapides, I. and Deutsch, Y., Mechanochemical adsorption of phenol by tot swelling clay minerals II simultaneous DTA and TG study. *J. Therm. Anal. Calorim.*, 1998, 51, 431–447.
6. Matteazi, P. and Le Caer, G., Room temperature mechanosynthesis of carbides by grinding of elemental powders. *J. Am. Ceram. Soc.*, 1991, 74, 1382–1390.
7. Alcala, M. D., Gotor, F. J., Perez-Maqueda, L. A., Real, C., Dianez, M. J. and Criado, J. M., Constant rate thermal analysis (CRTA) as a tool for the synthesis of materials with controlled texture and structure. *J. Therm. Anal. Calorim.*, 1999, 56, 1447–1452.
8. Grim, R. E., *Clay Mineralogy*. McGraw-Hill, New York, 1968.
9. Pascual, J., Zapatero, J., de Haro, M. C. J., Varona, I., Justo, A., Perez-Rodriguez, J. L. and Sanchez-Soto, P. J., Porous mullite and mullite-based composites by chemical processing of kaolinite and aluminium metal wastes. *J. Mater. Chem.*, 2000, 10, 1409–1414.
10. Miller, J. G. and Oulton, T. D., Prototropy in kaolinite during percussive grinding. *Clays Clay Miner.*, 1970, 18, 313–323.
11. Cicel, B. and Kranz, G., Mechanism of montmorillonite structure degradation by percussive grinding. *Clay Miner.*, 1981, 16, 151–162.
12. Juhasz, Z. and Somogoy, A., Grinding test with illite. *Keram.*, 1984, 36, 659–662.
13. Peŕez-Rodrı́guez, J. L., Madrid, L. and Sańchez Soto, P. J., Effects of dry grinding on pyrophyllite. *Clay Miner.*, 1988, 23, 399–410.
14. Sanchez-Soto, P. J., Wiewiora, A., Aviles, M. A., Justo, A., Perez-Maqueda, L. A., Perez-Rodriguez, J. L. and Bylina, P., Talc from Puebla de Lillo, Spain. II. Effect of dry grinding on particle size and shape. *Appl. Clay Sci.*, 1997, 12, 297–312.
15. Sanchez-Soto, P. J., Ruiz-Conde, A., Aviles, M. A., Justo, A. and Peŕez-Rodrı́guez, J. L., Mechanochemical effects on vermiculite and its influence on the synthesis of nitrogen ceramics. In *Ceramic Charting the Future*, ed. P. Vicenzeni. Techno Srl, Faenza (Italy), 1995, pp. 1383–1390.
16. Sanchez-Soto, P. J., Del Carmen Jimenez de Haro, M., Perez-Maqueda, L. A., Varona, I. and Perez-Rodriguez, J. L., Effects of dry grinding on the structural changes of kaolinite powders. *J. Am. Ceram. Soc.*, 2000, 83, 1649–1657.
17. Perez-Maqueda, L. A., Caneo, O. B., Poyato, J. and Perez-Rodriguez, J. L., Preparation and characterization of micron and submicron-sized vermiculite. *Phys. Chem. Miner.*, 2001, 28, 61–66.
18. Suslick, K. S., Application of ultrasound to materials chemistry. *MRS Bull.*, 1996, 74, 29–34.
19. Perez-Rodriguez, J. L., Carrera, F., Poyato, J. and Perez-Maqueda, L. A., Sonication as a tool for preparing nanometric vermiculite particles. *Nanotechnology*, 2002, 13, 382–387.
20. Wiewiora, A., Perez-Rodriguez, J. L., Perez-Maqueda, L. A. and Drapala, J., Particle size distribution in sonicated high- and low-charge vermiculites. *Appl. Clay Sci.* (in press).
21. Kupar, P. C., Kinetics of solid-state reactions of particulate ensembles with size distributions. *J. Am. Ceram. Soc.*, 1973, 56, 79–81.
22. Koga, N. and Criado, J. M., Kinetic analyses of solid-state reactions with a particle-size distribution. *J. Am. Ceram. Soc.*, 1998, 81, 2901–2909.
23. Lahiri, A. K., The effect of particle size distribution on TG. *Thermochimica Acta*, 1980, 40, 289–295.
24. Cooper, E. A. and Mason, T. O., Mechanism of $La_2CuO_4$ solid state powder reaction by quantitave XRD and impedance spectroscopy. *J. Am. Ceram. Soc.*, 1995, 78, 857–864.
25. Peŕez-Maqueda, L. A., Peŕez-Rodrı́guez, J. L., Scheiffele, G. W., Justo, A. and Sańchez-Soto, P. J., Thermal analysis of ground kaolinite and pyrophyllite. *J. Therm. Anal.*, 1993, 39, 1055–1067.
26. Peŕez-Rodrı́guez, J. L. and Sańchez-Soto, P. J., The influence of dry grinding on the thermal behaviour of pyrophyllite. *J. Therm. Anal.*, 1991, 37, 1401–1413.
27. Peŕez-Rodrı́guez, J., Peŕez-Maqueda, L., Justo, A. and Sańchez-Soto, P., Influence of grinding contamination on high temperature phases of kaolinite. *Ind. Ceram.*, 1992, 12, 109–113.
28. Wiewiora, A., Sanchez-Soto, P. J., Aviles, M. A., Justo, A., Perez-Maqueda, L. A., Perez-Rodriguez, J. L. and Bylina, P., Talc from Puebla de Lillo, Spain. I. XRD study. *Appl. Clay Sci.*, 1997, 12, 233–245.

29. Pérez-Maqueda, L. A., Franco, F., Avilés, M. A., Poyato, J. and Pérez-Rodríguez, J. L., Effect of sonication on particle-size distribution in natural muscovite and biotite. *Clays Clay Miner.*, 2003, 51, 701–708.
30. Jackson, M. L., *Soil Chemical Analysis—Advanced Course*, 2nd edn. Published by the author, Madison, WI, 1975.
31. Sanz, J. and Serratosa, J. M., Distinction of tetrahedrally and octahedrally coordinated Al in phyllosilicates by NMR-spectroscopy. *Clay Minerals*, 1984, 19, 113–115.
32. Sanz, J. and Serratosa, J. M., Si-29 and Al-27 high-resolution MAS–NMR spectra of phyllosilicates. *J. Am. Chem. Soc.*, 1984, 106, 4790–4793.
33. Wiewióra, A. and Weiss, Z., X-ray powder transmission diffractometry determination of mica polytypes: method and application to natural samples. *Clay Miner.*, 1985, 20, 231–248.
34. Mackenzie, R. C., Simple phyllosilicates based on gibbsite and brucite-like sheets. In *Differential Thermal Analysis, Vol. 1*, ed. R. Mackenzie. Academic Press, London, 1970, pp. 498–537.